# Magnetic-field assisted laser ablation of silicon

M. Schäfer[1,*], P. N. Terekhin[2], Y. Kang[1], G. Torosyan[1], X. del Arco Fargas[2], S. Hirtle[2], B. Rethfeld[2], and J. A. L'huillier[1]

[1]Photonik-Zentrum Kaiserslautern e.V. and Research Center OPTIMAS, Technische Universität Kaiserslautern, Kohlenhof Straße 10, Kaiserslautern 67663, Germany
[2]Department of Physics and Research Center OPTIMAS, Technische Universität Kaiserslautern, Erwin-Schrödinger-Straße 46, Kaiserslautern 67663, Germany

**Abstract**
Understanding and manipulation of the laser processing quality during the ablation of solids have crucial importance from fundamental and industrial perspectives. Here we have studied the effect of external magnetic field on the micro-material processing of silicon by ultrashort laser pulses. It was found experimentally that such a field directed along the laser beam improves the quality and efficiency of the material removal. Additionally, we observe that the formation of laser-induced periodic surface structures (LIPSS) in a multi-pulse regime is affected by the external magnetic field. Our results open a route towards efficient and controllable ultrafast laser micromachining.

*Corresponding author.
E-mail address: mareike.schaefer@pzkl.de

## 1. Introduction

Silicon is considered the most universal and broadly used material in the high technology industry due to its attractive physical, optical, chemical, and mechanical features [1,2]. Thus, it has already been investigated extensively in various research fields while still being studied for more potential applications. In the meantime, it has also developed into one of the main materials in laser micromachining. Presently, laser ablation of crystalline silicon is a well-studied and widely used processing technique. However, some of its aspects, for example, how the process is influenced by external electromagnetic fields, are still missing or need more detailed investigations. Assisted by an external field, the processes are expected to improve in the quality as well as efficiency. Revealing the optimal conditions for ablation on the surface and in volume processing of the material by means of pulsed laser beams is of very high interest. Changing the ambient atmosphere, including conditions, namely, the application of various processing gases [3] or liquid flows top layers [4-6], gives additional flexibility to the ablation process. Various studies have been carried out by some research groups to investigate the effect of the magnetic field on the ablation process as well [7-15]. However, the applied field in these studies was mainly perpendicular to the laser beam close to the surface. In our approach, an external magnetic field is applied axially along the laser beam close to the processed area. It is believed to generate a different effect during the ablation process, which is supposed to control it for a better outcome. Additionally, it is expected that a magnetic field allows a stronger absorption of laser pulses in silicon. Moreover, it shall provide local confinement for the hot electrons leading to enhanced deposition of the laser energy into the material.

To reach the most effective ways to process crystalline silicon, the parameters of laser beams and pulses still need to be optimized. This paper aims to investigate the influence of the external magnetic field on the ablation process following ultrashort pulses. The positive effect of the external magnetic field has already been demonstrated using nanosecond [10,11], picosecond [15], and femtosecond laser pulses [12-14]. In our case, femtosecond laser pulses in the infrared range have been applied to investigate the effects and driving mechanisms in the micromachining of silicon surfaces with the externally applied magnetic field.

## 2. Experimental setup

The effects of the externally applied magnetic field on laser treatment of the silicon surface have been studied using a femtosecond fiber laser system BlueCut from Menlo Systems GmbH (Germany). The system is able to deliver ultrashort pulses with a repetition rate from 10 kHz to 5 MHz (see Table 1) with a high beam quality $M^2<1.25$. The experiments of the present work have been mostly performed at the lowest possible repetition rate using 10 kHz. For deflecting and focusing the beam on the workpiece, a galvanometric scanner system (HurrySCANR©II/14, SCANLAB AG) with a telecentric F-theta optics (f≈100 mm, Linos respectively Jenoptik AG) was used. By measuring the laser beam spot size, the fit method from Liu [16] was applied. The profiling camera WinCamD (Laser2000) was applied systematically for checking the laser beam profile quality on its cross-section. The used laser parameters and the achieved spot size on the workpiece are summarized in Table 1.

Table 1. Laser parameters.

| Wavelength | Pulse duration | Pulse repetition rate | Maximum pulse energy | Minimum spot radius ($1/e^2$ of the maximum) |
|---|---|---|---|---|
| 1030 nm | 400 fs (FWHM) | 10 kHz | 10 µJ | 9,5 µm |

The magnetic field was created by using two identical permanent ring magnets with internal and external diameters of 30 mm and 70 mm, correspondingly. The magnets were arranged and held in the main aluminum holder one above the other in a common geometrical axis. The upper magnet was fixed in the main aluminum holder, while the bottom magnet could be moved up and down for smoothly varying in a wide range of the strength of the magnetic field created between the magnet rings. Point-to-point measurements of the magnetic flux density within that volume have been performed with a tesla meter device of a model FM 302 (Projekt Elektronik GmbH) by using its axial probe as a sensor. In our arrangement, the range from 50 mT to 230 mT could be covered and applied onto the working plane of silicon samples. Fig. 1(a) shows a cross-section of the configuration of the magnet with the simulated behavior of the magnetic field indicating lines. The magnetic field is parallel to the laser beam and perpendicular to the workpiece surface. A wide enough homogenous area of magnetic flux density is achieved in the center, as shown in Fig. 1(b). The magnetic flux density was measured over an area of approximately 14×14 mm$^2$ and demonstrated a homogeneous distribution over the central area with a diameter of about 9 mm. The latter was used for positioning the samples, which ensured the constant effect of the magnetic field for all ablation points that fall into this area. The samples used in our experiments were 10×10 mm$^2$ pieces cut from a p-type (boron-doped) silicon wafer with a standard thickness of 500 µm.

The ablated structures, including single-, multi-pulse and line ablation regimes, have been investigated and analyzed with a light microscope (Axio Imager.M2m, Carl Zeiss Microscopy

GmbH) using an optical circular difference interference contrast (C-DIC) mode. For more detailed structural measurements, a confocal microscope (Smartproof 5, Carl Zeiss Microscopy GmbH) with a vertical resolution of 1 nm and a scanning electron microscope (EVO-MA10, Carl Zeiss Microscopy GmbH) were applied.

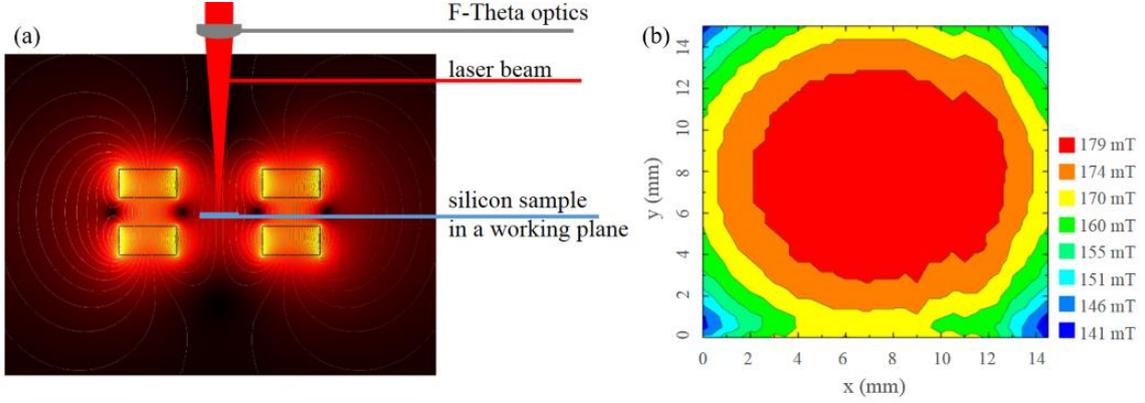

Fig. 1. (a) A cross-section at the arrangement of two ring magnets with simulated behavior of magnetic field lines and the position of the working plane in respect to the laser beam. (b) The behavior of the magnetic flux density in the center between the two ring magnets in dependence of space coordinates.

## 3. Results and discussion

The single-shot ablation process was achieved by using the scanner in the linear mode with the scanning speed of 2000 mm/s, so that the pulses were separated along the scanning line by a spot-to-spot distance of 200 μm. The temporal distance between spots corresponding to 10 kHz repetition rate is 100 μs. Measurements have been performed for different pulse energies with ($B$=150 mT, 200 mT) and without magnetic field ($B$=0 mT). Our experiments show that the samples of p-type crystalline silicon are modified in the presence of the external magnetic field, which leads to a better absorption (Landau-levels [17]) as well as more efficient gentle ablation. The measured ablated diameter with an applied magnetic field, $D$ ($B \neq 0$), was put into relation with that in the absence of a magnetic field, $D$ ($B=0$). The relative change of the crater diameter with the applied magnetic field is given by Eq. (1). A similar method was used to the change of ablation depth $\Delta h$.

$$\Delta D = \frac{D(B \neq 0) - D(B = 0)}{D(B = 0)}. \quad (1)$$

Since the morphology of the single-pulse ablated craters varies in different fluence ranges, the results are discussed in two regimes: low fluences (0 - 2.4 J/cm$^2$) and high fluences (above 2.4 J/cm$^2$). The effect of the external magnetic field also emphasizes different aspects in these two fluence regimes. In the low fluence regime, the external magnetic field mainly influences the absorption of the laser energy via the generated Landau levels. In contrast, at high laser fluences, the generated plasma after the laser pulse affects the dynamics of the ablation process. We show in Fig. 2 the relative change of the crater diameter as well as the removal depth as a function of the multiple of the damage threshold fluence. The maximum relative change of the ablation diameter takes place at the first detected ablation points processed close to the damage threshold fluence, which was determined as $F_{d\text{-th}}^{B=0}$=0.66 ± 0.01 J/cm$^2$ using the Liu method [16]. By increasing the applied fluence, the effect of the external magnetic field drops to the case without external magnetic field application at around 2.1$F_{d-th}$ for *B*=150 mT as well as *B*=200 mT. In Fig. 2(b), we can observe that an application of the magnetic field also produces a noticeable increase in the removal depth near the damage threshold fluence. At the field strength *B*=200 mT the removal depth of the craters is up to 50% deeper around the $F_{d-th}$, while the relative change in the removal depth shows no more increase when the fluence is bigger than 1.2$F_{d-th}$. At *B*=150 mT assisted ablation, there is an enhancement on the removal depth of approximately 10% at the first detected point. However, the removal depth becomes 5% - 25% smaller than the craters ablated without magnetic field with the further increase of the pulse fluence. Therefore, we can conclude that the strongest enhancement of the ablation due to the magnetic field is observed at fluences close to the single-pulse damage threshold fluence. However, in the high fluence regime, the dependence of the relative change of the damage diameter Δ*D* and the relative change of the removal depth Δ*h* on the multiple of the threshold fluence is complex and non-monotonic. Therefore, detailed explanations of these behaviors call for additional research, which is out of the scope of the current work.

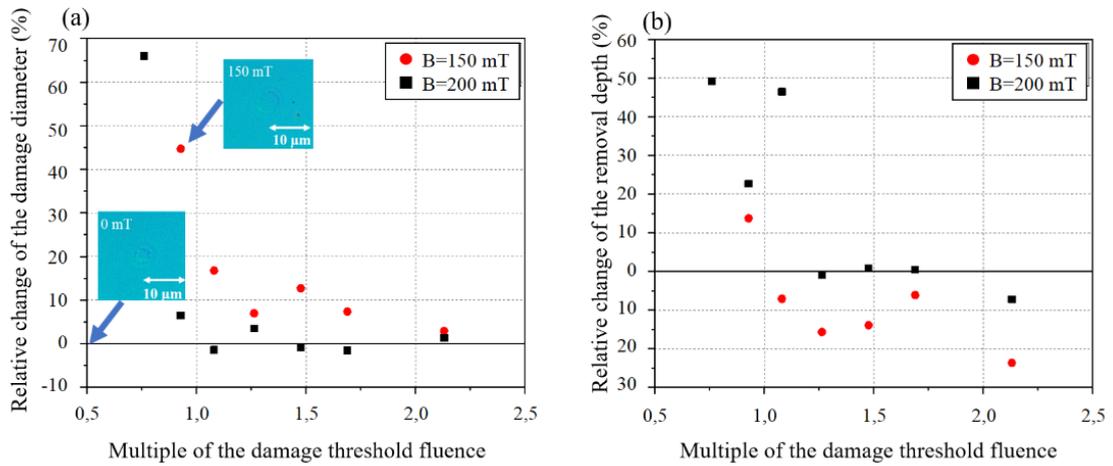

Fig. 2. (a) The relative change of the damage diameter *D* and (b) the relative change of the removal depth *h* in dependence of the multiple of the damage threshold fluence after femtosecond excitation of the silicon sample in the case of a single pulse N=1. The insets in Fig. 2(a) show the relative change of the damage diameter *D* at the laser fluence 0.93$F_{d-th}$ and the magnetic field strength of 0 mT and 150 mT.

A theoretical model in the frame of the Monte Carlo approach was developed to predict the influence of the external magnetic field on the lattice energy distribution after laser excitation. This model tracks the movement of individual electrons and considers scattering events. Elastic scattering describes the energy transfer to the lattice, impact ionization increases the number of traced electrons, and dummy collisions are implemented in order to keep the collision times flexible. The electrons are excited instantaneously by a gaussian laser profile in space. Following the excitation, a random direction of movement is assigned to the excited electrons. After a fixed interaction time, a scattering event takes place, in which the electron can transfer part of its energy to another particle. The direction of movement of the traced electron may also change. Our simulations show that the model assumptions strongly influence the spatial lattice energy distribution after the exchange of energy between the electron and lattice subsystems has taken place. Fig. 3 shows the spatial lattice energy distribution after the exchange of energy between electrons and lattice, where different assumptions were made. Namely, in Fig. 3(a) and 3(b), it was assumed that electrons do not change their direction of movement after an elastic scattering event. In contrast, in Fig. 3(c) and 3(d), the direction of movement is changed randomly.

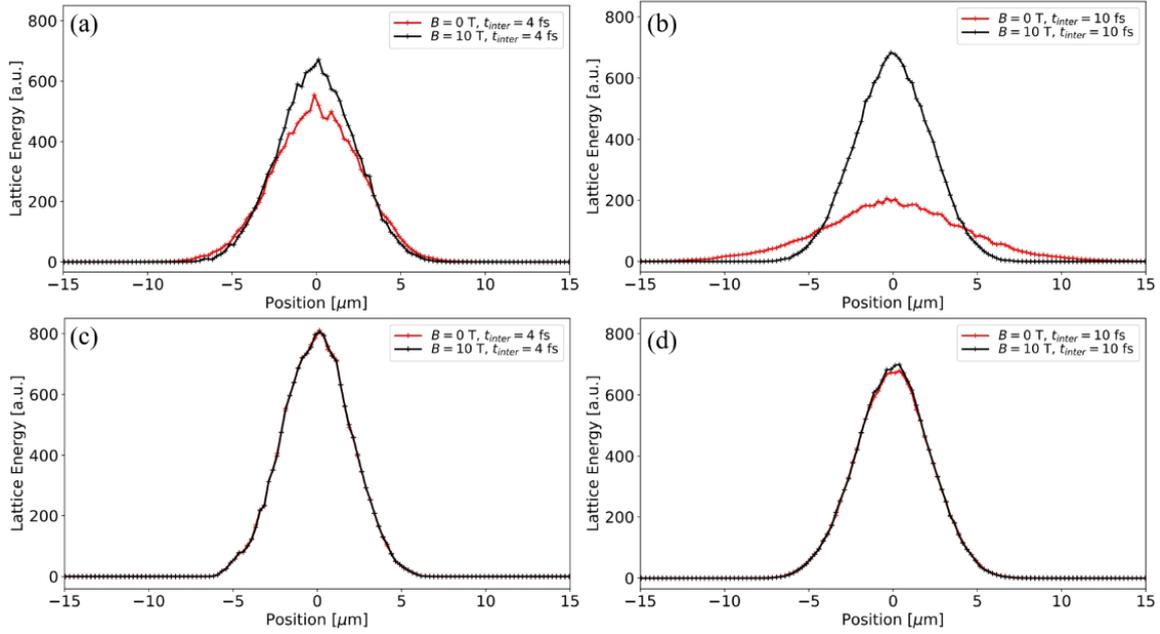

Fig. 3. Spatial distribution of the lattice energy after the energy exchange between the electron and lattice subsystems has occurred. It was assumed for (a) and (b) that electrons keep their direction of movement. In contrast, it was considered for (c) and (d) that electrons randomly change their direction of movement. Two fixed times between interactions were assumed: 4 fs and 10 fs.

We can observe in Fig. 3(a) and 3(b) that with the assumption of electrons keeping their direction of movement after an elastic scattering event, better spatial confinement of the lattice energy can be achieved with a magnetic field (black curves) compared to without magnetic field (red curves). At higher interaction times, this effect is pronounced clearer. This can be explained by the energy exchange between electrons and the lattice taking longer for higher interaction times due to less frequent electron-lattice collisions. Therefore, the electron distribution can diverge more without a magnetic field during the energy exchange, which leads to a more spread-out lattice energy distribution. With the assumption of electrons changing their direction of movement randomly, the scattering events prevent the electron distribution from spreading out. Due to this fact, the difference between the distributions with and without a magnetic field is smaller. However, the assumption of electrons not changing their direction of movement is a non-realistic assumption, such that the effects of the magnetic field are expected to be smaller in reality. This is supported by the experiments away from the damage threshold, where also only a small influence of the magnetic field was observed. However, at fluences close to the damage threshold, additional investigations are needed.

In the multi-pulse regime, 10-pulse ablation experiments were carried out on silicon samples. An ablation threshold fluence for 10-pulse irradiation without magnetic field $F_{th}$=0.29 J/cm² is almost twice lower than a single-pulse damage threshold. Due to the widely reported incubation effect, a non-ablating modification of the sample results in the lowering of the threshold by each pulse for the next one [18]. The results after 10-pulse ablation are presented in Fig. 4.

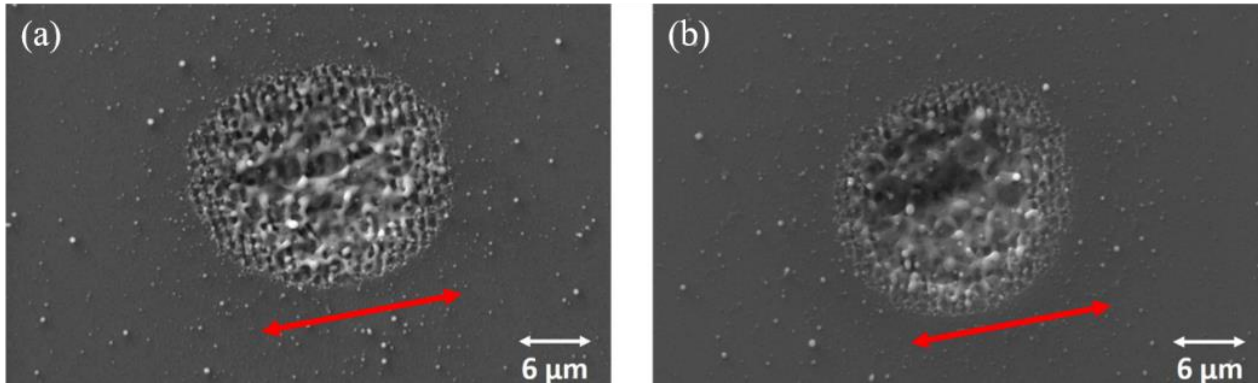

Fig. 4. SEM pictures of N=10 pulse ablated Si. (a) *B*=0 mT, (b) *B*=200 mT. The red arrows show the polarization direction of the incident laser beam with the structured lines oriented perpendicular to the polarization plane. The determined LIPSS periodicities are approximately 750 µm at *B*=0 mT and 650 µm at *B*=200 mT.

We can observe in Fig. 4 the formation of surface structures at the edge area of the laser spot. These structures were intensively studied and called laser-induced periodic surface structures (LIPSS). Since the discovery [19], they have been observed in many investigations and got very high popularity [20-22]. LIPSS occur and are being detected on metals, semiconductors, and dielectrics [20-29] after irradiation with a linear polarized laser beam strong enough to modify it. The structuring of the surface is performed in a one-step which makes it technologically very attractive. As an easily realized universal phenomenon, LIPSS have a wide range of applications, especially for surface functionalization [20-22]. However, a complete understanding of the mechanisms involved in the process of LIPSS formation is still missing. It resulted in various theories, which try to explain the LIPSS origin [21,24,25,30-35], for example, in the frameworks of the two-temperature model [24,27,28,31,36,37], a combined electromagnetic and compressible Navier-Stokes approach [34] or molecular dynamics-based simulations [5,6,33,37-39]. Therefore, being a quite spread phenomenon, they demand more detailed investigation in the sense of controllable and reproducible structuring of the surface covered by them. LIPSS may have periodicity varying in a wide range, depending on both the physical properties of the sample and laser parameters. Although in this paper we concentrate our attention mainly on the effect of the external magnetic field on the laser-induced ablation process on Si, we observe an influence of the external magnetic field also on LIPSS formation. Namely, the periodic structures were analyzed through the Fast Fourier Transformation and showed a single spatial frequency corresponding to the periodicity of 750 µm for *B*=0 mT and 650 µm for *B*=200 mT after 10 laser shots on Si. LIPSS are formed in our case due to the linearly polarized laser beam. We detect them with an orientation perpendicular to the polarization of the laser beam. As shown in Fig. 4, LIPSS are almost destroyed with the magnetic field, and the crater in the center of the ablation point got deeper. The hot electrons being generated by the fs-pulses are trapped more in the vicinity of the center. Hence the energy of the laser pulse is more transmitted into the depth of the sample and more efficiently used for ablation.

Further investigations on the effect of the external magnetic field on the ablation process have been performed on produced grooves by the laser pulse overlap of about 95%. As it is depicted in Fig. 5, the grooves without the applied field have an average width of about 6 µm with explicitly visible LIPSS along the grooves. As soon as the samples get surrounded by the magnetic field, the width of the grooves is increased almost twice to 12.1 µm with an applied field strength of 200 mT, and the LIPSS disappeared in this case. The destructing effect of external magnetic field on the LIPSS was observed in all our experiments in fs-regime and seemed to be general. Detailed investigation of LIPSS formation in the presence of the external magnetic field is left for our future work.

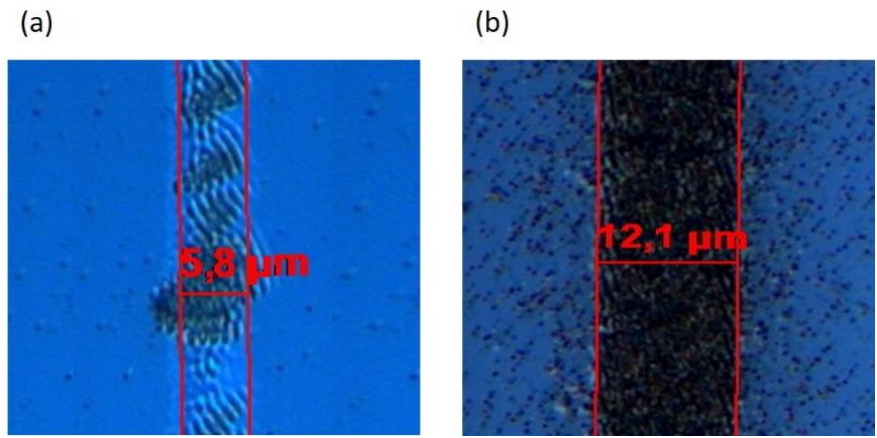

Fig. 5. Produced grooves following the fs-pulse scanning with 95% overlap
(a) B=0 mT and (b) B=200 mT.

## 4. Conclusions

We have shown that it is possible to control the quality of the laser microprocessing using the external magnetic field applied along the laser beam. The processed laser spot clearness and the depth, accessible with femtosecond laser pulses, demonstrate a positive effect. The strongest impact on the laser-induced modification of Si surfaces by an applied external magnetic field is observed at fluence levels close to the damage threshold. The results depend on the number of pulses applied at the processed points. Additionally, we observe that LIPSS formation is sensitive to the external magnetic field. Our results open a way for clean and efficient laser treatment of silicon with an opportunity to control the process of LIPSS formation externally by matching their properties to desired applications towards surface functionalization.

**Acknowledgments**

We acknowledge the financial support of the Federal Ministry of Education and Research (BMBF project "AssistAb" no. 13N14867 and 13N14868). We are grateful to the Applied University of Kaiserslautern, especially Prof. Dr.-Ing. Peter Starke, for using the confocal microscope. Simulations were executed on the high-performance cluster "Elwetritsch" through the projects TopNano and Mulan at the TU Kaiserslautern, which is a part of the "Alliance of High Performance Computing Rheinland-Pfalz". We kindly acknowledge the support of Regionales Hochschulrechenzentrum Kaiserslautern.